\documentclass[twocolumn]{jpsj3}
\usepackage{txfonts}
\RequirePackage{graphicx,color}

\title{
Theory of Photon-Assisted Magnetoacoustic Resonance \\
as a New Probe of Quadrupole Dynamics
}

\author{Mikito Koga$^1$ and Masashige Matsumoto$^2$}

\inst{
$^1$Department of Physics, Faculty of Education, Shizuoka University, Shizuoka
422--8529, Japan \\
$^2$Department of Physics, Faculty of Science, Shizuoka University, Shizuoka
422--8529, Japan \\
}

\abst{ 
Motivated by the recent progress of phonon-mediated control in quantum spin devices, we
propose a possibility of hybrid measurement using electron paramagnetic resonance (EPR)
and a surface acoustic wave (SAW).
Considering quadrupole-strain (QS) couplings suggested for silicon vacancies, we
present a minimum model of the two-level system to
investigate a magnetoacoustic resonance (MAR) coupled to various strain modes
driven by the SAW.
The longitudinal and transverse QS couplings can be changed by rotating a magnetic field,
which depends on a combination of the strain modes.
Using the Floquet theory, we elucidate each coupling effect on the time-averaged transition
probability, especially focus on a single-phonon transition process.
The important result is that the longitudinal QS coupling brings about a sharp photon-assisted
resonance and leads to an abrupt change in the field-angle dependent transition probability.
Since this phonon transition process is always accompanied by the photon transition, the field
angle for the sharp resonance peak can be detected by the EPR measurement.
The hybrid EPR-MAR measurement is useful to confirm the existence of quadrupole
degrees of freedom strongly coupled to elastic strains, and thus it is expected to be
a complementary probe for the precise evaluation of quadrupole properties.
}

\begin{document}

\maketitle

\newcommand{\ds}{\displaystyle}

\renewcommand{\H}{\mathcal{H}}
\newcommand{\br}{{\mbox{\boldmath$r$}}}
\newcommand{\bR}{{\mbox{\boldmath$R$}}}
\newcommand{\bS}{{\mbox{\boldmath$S$}}}
\newcommand{\bk}{{\mbox{\boldmath$k$}}}
\newcommand{\bH}{{\mbox{\boldmath$H$}}}
\newcommand{\bh}{{\mbox{\boldmath$h$}}}
\newcommand{\bJ}{{\mbox{\boldmath$J$}}}
\newcommand{\bI}{{\mbox{\boldmath$I$}}}
\newcommand{\bPsi}{{\mbox{\boldmath$\Psi$}}}
\newcommand{\bpsi}{{\mbox{\boldmath$\psi$}}}
\newcommand{\bPhi}{{\mbox{\boldmath$\Phi$}}}
\newcommand{\bd}{{\mbox{\boldmath$d$}}}
\newcommand{\bG}{{\mbox{\boldmath$G$}}}
\newcommand{\om}{{\omega_n}}
\newcommand{\omm}{{\omega_{n'}}}
\newcommand{\omd}{{\omega^2_n}}
\newcommand{\omt}{{\tilde{\omega}_{n}}}
\newcommand{\ommt}{{\tilde{\omega}_{n'}}}
\newcommand{\btau}{{\hat{\tau}}}
\newcommand{\brho}{{\mbox{\boldmath$\rho$}}}
\newcommand{\bsigma}{{\mbox{\boldmath$\sigma$}}}
\newcommand{\bSigma}{{\mbox{\boldmath$\Sigma$}}}
\newcommand{\bt}{{\hat{t}}}
\newcommand{\bq}{{\hat{q}}}
\newcommand{\bLambda}{{\hat{\Lambda}}}
\newcommand{\bDelta}{{\hat{\Delta}}}
\newcommand{\bu}{{\hat{u}}}
\newcommand{\bU}{{\hat{U}}}
\newcommand{\bskp}{{\mbox{\scriptsize\boldmath $k$}}}
\newcommand{\skp}{{\mbox{\scriptsize $k$}}}
\newcommand{\bsrp}{{\mbox{\scriptsize\boldmath $r$}}}
\newcommand{\bsRp}{{\mbox{\scriptsize\boldmath $R$}}}
\newcommand{\bsk}{\bskp}
\newcommand{\sk}{\skp}
\newcommand{\bsr}{\bsrp}
\newcommand{\bsR}{\bsRp}
\newcommand{\ri}{{\rm i}}
\newcommand{\re}{{\rm e}}
\newcommand{\rd}{{\rm d}}
\newcommand{\rM}{{\rm M}}
\newcommand{\rs}{{\rm s}}
\newcommand{\rt}{{\rm t}}
\newcommand{\Tc}{{$T_{\rm c}$}}
\renewcommand{\Pr}{{PrOs$_4$Sb$_{12}$}}
\newcommand{\La}{{LaOs$_4$Sb$_{12}$}}
\newcommand{\LaPr}{{(La$_{1-x}$Pr${_x}$)Os$_4$Sb$_{12}$}}
\newcommand{\PrLa}{{(Pr$_{1-x}$La${_x}$)Os$_4$Sb$_{12}$}}
\newcommand{\OsRu}{{Pr(Os$_{1-x}$Ru$_x$)$_4$Sb$_{12}$}}
\newcommand{\PrRu}{{PrRu$_4$Sb$_{12}$}}
\newcommand{\Haa}{{\mbox{\small $H_{-2}^{[0]}$}}}
\newcommand{\Hbb}{{\mbox{\small $H_{-1}^{[0]}$}}}
\newcommand{\Hcc}{{\mbox{\small $H_{-0}^{[0]}$}}}
\newcommand{\Hdd}{{\mbox{\small $H_{1}^{[0]}$}}}
\newcommand{\Hee}{{\mbox{\small $H_{2}^{[0]}$}}}
\newcommand{\Hab}{{\mbox{\small $H^{[-1]}$}}}
\newcommand{\Hba}{{\mbox{\small $H^{[1]}$}}}
\newcommand{\Hzero}{{\mbox{\small $\bf 0$}}}
\newcommand{\sddots}{{\mbox{\small $\ddots$}}}
\newcommand{\svdots}{{\mbox{\small $\vdots$}}}
\newcommand{\scdots}{{\mbox{\small $\cdots$}}}

\section{Introduction}
Quadrupole degrees of freedom are important factors as well as spin degrees of freedom for
high spin states possessing degenerate orbitals.
A possibility of electromechanical control of nuclear spins has been reported by the recent
experiment of a GaAs-based resonator.
\cite{Okazaki18}
The emergence of sideband nuclear magnetic resonance peaks can be explained by the
effect of dynamical quadrupole-strain (QS) coupling on the nuclear spin state.
There is also a proposal of mechanically and electrically driven electron spin resonance for
nitrogen-vacancy (NV) centers in diamond.
\cite{Udvarhelyi18}
Experimentally, for the NV centers, phonon induced orbital transitions were detected by
photoluminescence excitation spectroscopy.
\cite{Golter16a,Golter16b,Chen18}
Motivated by such progress of phonon-mediated control in spintronics, we theoretically investigate
a possibility of hybrid measurement using electron paramagnetic resonance (EPR) and a surface
acoustic wave (SAW).
\cite{Sasaki19}
For solid-state electrons, the dynamical QS coupling can be driven by the SAW.
By combination of photon and phonon transition processes, more precise analysis is expected
to identify the symmetries of quadrupoles and also to reveal hidden quadrupole properties.
We propose that the photon-assisted magnetoacoustic resonance (MAR) can be adopted as a
complementary or an alternative probe to the conventional measurements of quadrupole such as
elastic softening, nuclear quadrupole resonance, resonant x-ray diffraction, and polarized neutron
diffraction.
\cite{Kuramoto09}
\par

In particular, we can apply the idea of the hybrid EPR-MAR measurement to
confirm an extremely strong strain coupling with a quadrupole moment that possibly emerges in a
silicon vacancy.
This surprisingly strong QS coupling was first reported by bulk measurements of elastic softening
in a boron-doped silicon wafer.
\cite{Goto06,Baba13,Okabe13,Mitsumoto14}
The significance of an orbitally degenerate vacancy state was also suggested by early
theoretical studies.
\cite{Matsuura08,Yamada09,Ogawa11}
The precise evaluation of vacancy concentration in the surface of a silicon wafer is inevitably
required for information technology of semiconductor devices.
In the SAW measurement for the silicon vacancy, the elastic stains are driven by oscillating strain
modes such as selected quadrupole symmetries $2z^2 - x^2 - y^2$, $x^2 - y^2$, and $zx$ for the
SAW propagating in the $x$ direction of the crystal axis.
\cite{Mitsumoto14}
Owing to the magnetic-field-dependent multiplet,
each QS coupling can be changed by rotating an applied magnetic field and can be evaluated by
probing a coupled photon-phonon transition in the EPR measurement.
\par

The Floquet theory is very powerful for examining the two-level system that interacts with a
periodically oscillating field.
In the Floquet formalism, a problem of solving a time-dependent Schr$\ddot{\rm o}$dinger
equation is transformed to a time-independent eigenvalue problem based on an
infinite-dimensional matrix form of the Floquet Hamiltonian.
\cite{Shirley65,Son09,Hausinger10,Eckardt15,Mikami16}
In the present study, there are both diagonal (longitudinal) and off-diagonal
(transverse) couplings with the oscillating field related to various phonon modes.
It is the most important to elucidate the former coupling effect on the photon-assisted
transition in a single-phonon resonance process.
The significance of the longitudinal coupling effect was previously studied in a different context of
a superconducting quantum interference device driven by an ac field.
\cite{Son09,Hausinger10}
The key points are the energy shift of a sharp resonance peak and the peak broadening.
Considering the transverse coupling effect as well, we extend the Floquet theory, which was
frequently applied to multiphoton resonance processes in quantum devices,
\cite{Chu04}
to investigate a more complicated problem of phonon-mediated transition processes in solid-state
electronic spin systems.
\par

This paper is organized as follows.
In the first part of Sect.~2, we present a minimum model for the two-level system including
QS couplings and derive this model from a realistic system suggested for the silicon vacancy.
We compute the longitudinal and transverse QS couplings with the strain modes, 
which are given as a function of the rotation angle of the magnetic field.
In the second part, the Floquet Hamiltonian is derived for the two-level system coupled to the
SAW phonons.
For nearly degenerate Floquet states, it is useful to analyze an effective $2 \times 2$ matrix
formulated by the Van Vleck perturbation theory for a multi-phonon resonance.
\cite{Son09,Hausinger10}
In Sect.~3, we elucidate the longitudinal-mode and transverse-mode coupling effects on the
time-averaged transition probability.
Next, we show the field-angle dependence of the transition probability and how to evaluate
the QS couplings with different symmetries.
Finally, the conclusion and discussion are given by Sect.~4. 

\section{Model}
\subsection{Quadrupole-strain couplings in two-level system}
\begin{figure}
\begin{center}
\includegraphics[width=7cm,clip]{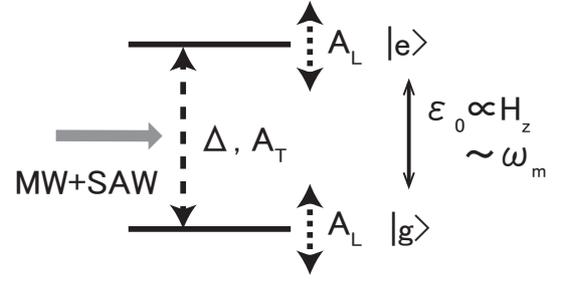}
\end{center}
\caption{
Two-level system ($| g \rangle$ and $| e \rangle$) coupled to strain modes driven by the SAW.
The transition between $| g \rangle$ and $| e \rangle$ occurs through the transverse QS coupling
($| A_T |$) in addition to the MW photon coupling ($\Delta$).
The transition is also due to the longitudinal QS coupling ($| A_L |$) in a photon-assisted
transition process.
Here, the detuning energy $\varepsilon_0$ is almost proportional to the magnetic field $H_z$
for the Zeeman splitting comparable to the single-phonon energy $\hbar \omega_m$
}
\label{fig:1}
\end{figure}

To extract essential roles of QS couplings in the transition probabilities detected
by the EPR, we study a minimum model of the two-level system described by the following
Hamiltonian $H = H_0 + H_{\rm QS} + H_{\rm MW}$:
\cite{Chen18}
\begin{align}
& H_0 = \frac{1}{2} \hbar \omega_0  ( - | g \rangle \langle g | + | e \rangle \langle e | ), \\
& H_{\rm QS} = \frac{1}{2} g_L \varepsilon_L (t) ( | g \rangle \langle g | - | e \rangle \langle e | )
\nonumber \\
&~~~~~~
+ \frac{1}{2} g_T \varepsilon_T (t) ( | g \rangle \langle e | + | e \rangle \langle g | ),
\label{eqn:HQS} \\
& H_{\rm MW} = \frac{1}{2} \hbar \Omega ( | g \rangle \langle e | e^{i \omega_l t}
+  | e \rangle \langle g | e^{ - i \omega_l t} ),
\label{eqn:HMW}
\end{align}
where $| g \rangle$ and $| e \rangle$ represent the wave functions of the ground and excited
states of the two-level system, respectively, with the energy difference $\hbar \omega_0$.
In Eq.~(\ref{eqn:HQS}), $g_L$ ($g_T$) is a QS coupling constant for the
longitudinal (transverse) vibration modes between the two states, and the displacement
$\varepsilon_L$ ($\varepsilon_T$) with the time $t$ dependence represents an elastic strain of
the acoustic-wave $L$-mode ($T$-mode) coupled to quadrupoles.
Equation~(\ref{eqn:HMW}) describes the absorption ($| e \rangle \langle g |$) and emission
($| g \rangle \langle e |$) processes of a microwave
(MW) with frequency $\omega_l$, and the Rabi frequency $\Omega$ characterizes the periodic
oscillation of the time-dependent transition probability.
Here, the rotating wave approximation (RWA) has been considered.
\cite{Chen18}
In the conventional manner, we introduce a unitary transformation $U = e^{ - i \omega_l I_z t}$,
where the pseudospin operator $\bI$ is defined by
\begin{align}
& I_z = \frac{1}{2} ( | g \rangle \langle g | - | e \rangle \langle e | ), \\
& I_x = \frac{1}{2} ( | g \rangle \langle e | + | e \rangle \langle g | ) \equiv \frac{1}{2} ( I_+ + I_- ),
\end{align}
and $I_y = i [ I_z, I_x]$.
Transforming the original Hamiltonian $H$ into the rotating flame, we have
\begin{align}
& \tilde{H} = U H U^\dagger - i \hbar U \frac{ \partial U^\dagger }{ \partial t }
\nonumber \\
&~~~
= - [ \hbar ( \omega_0 - \omega_l ) + g_L \varepsilon_L (t) ] I_z
\nonumber \\
&~~~~~~
+ \frac{1}{2} g_T \varepsilon_T (t)
( I_+ e^{ - i \omega_l t } +  I_- e^{ i \omega_l t} ) + \hbar \Omega I_x.
\end{align}
This means that the original Schr$\ddot{\rm o}$dinger equation is transformed as
\begin{align}
i \hbar \frac{ \partial }{ \partial t } | \psi \rangle = H | \psi \rangle~~\rightarrow~~
i \hbar \frac{ \partial }{ \partial t } U | \psi \rangle = \tilde{H} U | \psi \rangle.
\end{align}
For the propagating SAW, the elastic strain components are given by
\begin{align}
\varepsilon_L (t) = a_L \cos ( \omega_m t + \phi_L),~~
\varepsilon_T (t) = a_T \cos ( \omega_m t + \phi_T),
\end{align}
where $\omega_m$ is the SAW frequency, $a_L$ ($a_T$) is the amplitude of the
$L$-mode ($T$-mode), and the phase shift $\phi_L$ ($\phi_T$) depends on the positions in
the direction of the propagating SAW.
Notice that the phase difference $\phi_T - \phi_L$ is constant.
Keeping in mind an ultrasonic measurement of gigahertz order, we assume
$\omega_m \gg \omega_l$ ($\omega_m \pm \omega_l \simeq \omega_m$)  in the following
discussion.
For the high frequency SAW, the transformed Hamiltonian $\tilde{H}$ is reduced to
\begin{align}
& \tilde{H}(t) = - [ \varepsilon_0 + \frac{1}{2} ( A_L e^{ i \omega_m t } + A_L^* e^{ - i \omega_m t } ) ]
I_z
\nonumber \\
&~~~~~~
+ \frac{1}{2} ( A_T e^{ i \omega_m t} + A_T^* e^{ - i \omega_m t} ) I_x
\nonumber \\
&~~~~~~
+ \frac{1}{2} ( \Delta^* I_+ + \Delta I_- ),
\label{eqn:Ht}
\end{align}
where
\begin{align}
\varepsilon_0 = \hbar ( \omega_0 - \omega_l ),~~A_L = - g_L a_L e^{i \phi_L},~~
A_T = g_T a_T e^{i \phi_T}.
\label{eqn:parameter}
\end{align}
In the last term of $\tilde{H}(t)$, we consider a relative phase difference of the coupled photon
and phonon as $\Delta = | \Delta | e^{i \phi_l} = \hbar \Omega e^{i \phi_l}$, and $\phi_l$ may take
arbitrary values owing to random distributions of local quadrupoles (vacancies) in a crystal.
For $| \Delta | / \hbar \omega_m \ll 1$, the calculated transition probability is not much dependent
on $\phi_l$ as discussed later.
Each term in Eq.~(\ref{eqn:Ht}) is sketched by Fig.~\ref{fig:1}.
For understanding the present two-level system, it is helpful to express a matrix
form of $\tilde{H}(t)$ on the basis of $\{ | g \rangle, | e \rangle \}$ as
\begin{align}
\tilde{H}(t) = \frac{1}{2}
\left(
\begin{array}{cc}
- \varepsilon_0 & \Delta^* \\
\Delta & \varepsilon_0
\end{array}
\right)
+ \frac{1}{2}
\left(
\begin{array}{cc}
- A_L ( \omega_m t ) & A_T ( \omega_m t ) \\
A_T ( \omega_m t ) & A_L ( \omega_m t )
\end{array}
\right),
\end{align}
and $A_{L(T)} ( \omega_m t ) = ( A_{L(T)} e^{i \omega_m t } + A_{L(T)}^* e^{- i \omega_m t} ) / 2$.
For $\Delta = 0$, the time-dependent diagonal coupling with $A_L$ only changes the energy
difference $\varepsilon_0 = \hbar \omega_0$ and does not contribute to the direct transition
between $| g \rangle$ and $| e \rangle$.
However, this coupling plays an important role in a photon-assisted transition process for a finite
$\Delta$.
\par

\subsection{Derivation of two-level system in a realistic case}
Here, considering the quartet ground state proposed for the silicon vacancy, we describe the
QS interaction with elastic strains driven by the SAW.
A combination of QS couplings with different symmetries is changed by rotating an applied
magnetic field.
We derive the two-level system for the ground and first excited states lifted from the degenerate
quartet in the magnetic field.
The derivation presented here is also applicable to other quartet systems as well as the silicon
vacancy.
\par

\begin{figure}
\begin{center}
\includegraphics[width=7cm,clip]{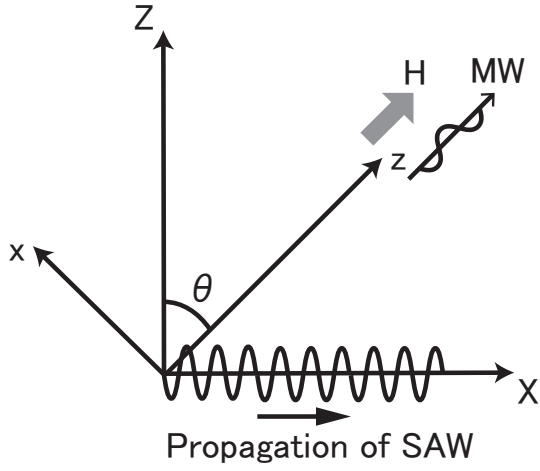}
\end{center}
\caption{
Schematic diagram of hybrid measurement with EPR and SAW.
The $(X, Z)$ and $(x,z)$ coordinates are set for the SAW propagating along the $X$ axis and for
the rotating static magnetic field $H$ parallel to the $z$ axis, respectively.
The MW magnetic field oscillates along the $x$ axis.
For a silicon wafer, vacancies are distributed randomly in the XY plane of the surface layer.
}
\label{fig:2}
\end{figure}
In the boron-doped silicon wafer, the elastic softening is well explained by
the strong QS coupling with the vacancy orbital.
\cite{Baba13,Okabe13}
It is expected that the electronic state with orbital $L = 1$ and spin $S = 1/2$ forms
the $\Gamma_8$ quartet ground and $\Gamma_7$ doublet excited states, owing to the
spin-orbit interaction in the $T_d$ cubic crystal-field environment.
We restrict ourselves to the $\Gamma_8$ quartet described by the
$J = 3/2$ angular momentum operator.
As illustrated in Fig.~\ref{fig:2}, we consider that the SAW propagates in the $X || [100]$ direction
on the $Z || [001]$ surface of the silicon wafer in the presence of an external magnetic field rotating
around $Y || [010]$ in the $ZX$ plane.
Under this experimental setting, the elastic stains coupled to the quadrupoles are given
by the following three components:
\cite{Mitsumoto14}
\begin{align}
& \varepsilon_U = \frac{1}{\sqrt{3}} ( 2 \varepsilon_{ZZ} - \varepsilon_{XX} - \varepsilon_{YY} ), \\
& \varepsilon_V = \varepsilon_{XX} - \varepsilon_{YY}, \\
& \varepsilon_{ZX} = \frac{\partial u_X}{\partial Z} + \frac{\partial u_Z}{\partial X},
\end{align}
where $u_i$ ($i = X, Y, Z$) is a displacement vector component and
$\varepsilon_{ii} = \partial u_i / \partial x_i$ ($x_i = i = X, Y, Z$).
Here, an irrelevant $u_Y$ has been considered for the SAW propagating along the $X$ axis, and
$\varepsilon_{YZ} = \varepsilon_{XY} = 0$.
\cite{Mitsumoto14}
As a consequence, the interaction of the $\Gamma_8$ quadrupoles and local strains in the
magnetic field can be described by $H_{\rm local} = H_{\rm Zeeman} + H_{\rm QS}$:
\begin{align}
& H_{\rm Zeeman} = - g_J \mu_{\rm B} \bJ \cdot \bH,
\label{eqn:HZ} \\
& H_{\rm QS} = g_{\Gamma_3} ( O_U \varepsilon_U + O_V \varepsilon_V )
+ g_{\Gamma_5} O_{ZX} \varepsilon_{ZX}.
\label{eqn:HQSg}
\end{align}
Here, the quadrupole operators are represented by the second-rank tensor operators of
$\bJ = (J_X, J_Y, J_Z)$ as
\begin{align}
& O_U = \frac{1}{\sqrt{3}} ( 2 J_Z^2 - J_X^2 - J_Y^2 ),
\label{eqn:OU} \\
& O_V = J_X^2 - J_Y^2, \\
& O_{ZX} = J_Z J_X + J_X J_Z.
\label{eqn:OZX}
\end{align}
In Eq.~(\ref{eqn:HZ}) of the Zeeman term, $g_J$ is the Land$\acute{\rm e}$'s $g$ factor and
$\mu_{\rm B}$ is the Bohr magneton.
Equation (\ref{eqn:HQSg}) consists of the $\Gamma_3$ and $\Gamma_5$ irreducible
representations in the cubic point group with the QS coupling constants
$g_{\Gamma_3}$ and $g_{\Gamma_5}$, respectively.
\par

Next, we consider the rotation of the magnetic field and choose the $z$ axis parallel to
$\bH / |\bH| = (\sin \theta, 0, \cos \theta)$ in the $ZX$ plane.
Accordingly, the $x$ and $y$ axes are set as $x || ( - \cos \theta, 0, \sin \theta )$ and
$y || ( 0, -1, 0)$. 
The components of the $\bJ$ operator in the $(xyz)$ coordinate are related to those in the
$(XYZ)$ crystal coordinate as
\begin{align}
\left\{
\begin{array}{l}
J_x = - \cos \theta \cdot J_X + \sin \theta \cdot J_Z, \\
J_y = - J_Y, \\
J_z = \sin \theta \cdot J_X + \cos \theta \cdot J_Z.
\end{array}
\right.
\end{align}
Under the field parallel to the $z$ axis, it is convenient to replace
the quadrupole operators in Eqs.~(\ref{eqn:OU})--(\ref{eqn:OZX}) by the linear combination of
$O_u = ( 2 J_z^2 - J_x^2 - J_y^2 ) / \sqrt{3}$, $O_v = J_x^2 - J_y^2$, and
$O_{zx} = J_z J_x + J_x J_z$:
\begin{align}
\left\{
\begin{array}{l}
O_U \rightarrow \ds{ - \frac{ 1 - 3 \cos^2 \theta }{2} O_u + \frac{\sqrt{3}}{2} \sin^2 \theta \cdot O_v } \\ ~~~~~~~~~~
+ \sqrt{3} \sin \theta \cos \theta \cdot O_{zx}, \\
O_V \rightarrow \ds{ \frac{\sqrt{3}}{2} \sin^2 \theta \cdot O_u + \frac{ 1 + \cos^2 \theta }{2} O_v }
- \sin \theta \cos \theta \cdot O_{zx}, \\
O_{ZX} \rightarrow \sqrt{3} \sin \theta \cos \theta \cdot O_u  -  \sin \theta \cos \theta \cdot O_v
- \cos 2 \theta \cdot O_{zx}.
\end{array}
\right.
\end{align}
On the basis of the eigenstates of $J_z$ ($= \pm 1/2, \pm 3/2$), the field-angle dependence of the QS interaction in Eq.~(\ref{eqn:HQSg}) is explicitly written as
$H_{\rm QS} = A_u O_u + A_v O_v + A_{zx} O_{zx}$, where
\begin{align}
& A_u = g_{\Gamma_3} \left[ \left( \frac{1}{4} \varepsilon_U + \frac{\sqrt{3}}{4} \varepsilon_V \right)
+ \left( \frac{3}{4} \varepsilon_U - \frac{\sqrt{3}}{4} \varepsilon_V \right) \cos 2 \theta \right]
\nonumber \\
&~~~~~~
+ g_{\Gamma_5} \left( \frac{\sqrt{3}}{2} \varepsilon_{ZX} \right) \sin 2 \theta, \\
& A_v = g_{\Gamma_3} \left[ \left( \frac{\sqrt{3}}{4} \varepsilon_U + \frac{3}{4} \varepsilon_V \right)
+ \left( - \frac{\sqrt{3}}{4} \varepsilon_U + \frac{1}{4} \varepsilon_V \right) \cos 2 \theta \right]
\nonumber \\
&~~~~~~
+ g_{\Gamma_5} \left( - \frac{1}{2} \varepsilon_{ZX} \right) \sin 2 \theta, \\
& A_{zx} = g_{\Gamma_3} \left( \frac{\sqrt{3}}{2} \varepsilon_U - \frac{1}{2} \varepsilon_V \right)
\sin 2 \theta
+ g_{\Gamma_5} \left( - \varepsilon_{ZX} \right) \cos 2 \theta. 
\end{align}
\par

For a strong magnetic field $g_J \mu_B H_z \sim \hbar \omega_m$, we focus on the
quadrupole-stain coupling between the ground state with $J_z = 3/2$ denoted by $| g \rangle$
and the first excited state with $J_z = 1/2$ denoted by $| e \rangle$ for $H_{\rm Zeeman}$.
In the subspace of the two states, $\hbar \omega_0 = g_J \mu_{\rm B} H_z$ in
Eq.~(\ref{eqn:parameter}), and the quadrupole operators $O_u$ and $O_{zx}$ are
reduced to
\begin{align}
& O_u = \sqrt{3} ( | g \rangle \langle g | - | e \rangle \langle e | ) = 2 \sqrt{3} I_z,
\label{eqn:Ou2} \\
& O_{zx} = \sqrt{3} ( | g \rangle \langle e | + | e \rangle \langle g | ) = 2 \sqrt{3} I_x,
\label{eqn:Ozx2}
\end{align}
and there is no $O_v$ coupling between $| g \rangle$ and $| e \rangle$.
For the local stain components of the propagating SAW,
\begin{align}
\varepsilon_i (t) = a_i \cos ( \omega_m t  + \phi_i )
\label{eqn:straini}
\end{align}
represents one of the three strain modes ($i = U, V, ZX$), where $a_i$ is the local amplitude of
each mode and the phase shift $\phi_i$ is given as $(- k X + \phi_{i,0})$ in the $X$
direction of the SAW with wave number $k$ ($= \omega_m / v$: $v$ is the sound 
velocity) and initial phase shift $\phi_{i,0}$.
Notice that the phase differences $\phi_i - \phi_j$ ($i, j = U, V, ZX$) are constant.
Since $A_u O_u$ and $A_{zx} O_{zx}$ correspond to the $L$-mode and $T$-mode in
Eq.~(\ref{eqn:Ht}), respectively, we obtain
\begin{align}
H_{\rm QS} = g_L a_L \cos ( \omega_m t + \phi_L ) I_z
+ g_T a_T \cos ( \omega_m t + \phi_T ) I_x,
\label{eqn:HQSLT}
\end{align}
where $g_\mu a_\mu$ ($\mu = L, T$) are determined from
\begin{align}
& g_\mu a_\mu
\left(
\begin{array}{l}
\cos \phi_\mu \\
\sin \phi_\mu
\end{array}
\right)
= C_{\mu,0} + C_{\mu,1} \cos 2 \theta + C_{\mu,2} \sin 2 \theta.
\label{eqn:ga}
\end{align}
The coefficients $C_{\mu,i}$ ($i = 0, 1, 2$) depend on $a_i$ and $\phi_i$ in
Eq.~(\ref{eqn:straini}) as
\begin{align}
& C_{L,0} = \sqrt{3} g_{\Gamma_3} \left[ \frac{1}{2} a_U
\left(
\begin{array}{l}
\cos \phi_U \\
\sin \phi_V
\end{array}
\right)
+ \frac{\sqrt{3}}{2} a_V
\left(
\begin{array}{l}
\cos \phi_V \\
\sin \phi_V
\end{array}
\right)
\right], \\
& C_{L,1} = 3 g_{\Gamma_3} \left[ \frac{\sqrt{3}}{2} a_U
\left(
\begin{array}{l}
\cos \phi_U \\
\sin \phi_U
\end{array}
\right)
- \frac{1}{2} a_V
\left(
\begin{array}{l}
\cos \phi_V \\
\sin \phi_V
\end{array}
\right)
\right], \\
& C_{L,2} = 3 g_{\Gamma_5} a_{ZX}
\left(
\begin{array}{l}
\cos \phi_{ZX} \\
\sin \phi_{ZX}
\end{array}
\right), \\
& C_{T,1} = - 2 \sqrt{3} g_{\Gamma_5} a_{ZX}
\left(
\begin{array}{l}
\cos \phi_{ZX} \\
\sin \phi_{ZX}
\end{array}
\right), \\
& C_{T,2} = 2 \sqrt{3} g_{\Gamma_3} \left[ \frac{\sqrt{3}}{2} a_U
\left(
\begin{array}{l}
\cos \phi_U \\
\sin \phi_U
\end{array}
\right)
- \frac{1}{2} a_V
\left(
\begin{array}{l}
\cos \phi_V \\
\sin \phi_V
\end{array}
\right)
\right],
\label{eqn:CT2}
\end{align}
and $C_{T,0}$ vanishes.
\par

\subsection{Floquet Hamiltonian}
According to the Floquet theory,
\cite{Shirley65}
the time-dependent Schr$\ddot{\rm o}$dinger equation
\begin{align}
i \hbar \frac{ \partial \psi(t) }{ \partial t } = H(t) \psi(t),
\end{align}
where $H (t)$ is periodic in time, is solved by the following eigenvalue equation
\begin{align}
\left[ H(t) - i \hbar \frac{ \partial }{ \partial t } \right] \phi(t) = q \phi(t).
\label{eqn:eigenvalue}
\end{align}
The time-periodic wave function $\phi(t)$ is related to $\psi(t)$ as
$\psi(t) = e^{-i q t / \hbar} \phi(t)$ with the quasienergy $q$.
The periodic time-dependent differential equation is transformed to a time-independent
infinite-dimensional matrix eigenvalue problem through the Fourier expansion
\begin{align}
H(t) = \sum_{n = -\infty}^{\infty} H^{[n]} e^{i n \omega t},~~
\phi(t) = \sum_{n = -\infty}^{\infty} \phi^{[n]} e^{i n \omega t},
\end{align}
and Eq.~(\ref{eqn:eigenvalue}) is reduced to
\begin{align}
\sum_m [ H^{[ n - m]} + n \hbar \omega \delta_{nm} ] \phi^{[m]} = q \phi^{[n]},
\label{eqn:Hq}
\end{align}
where $\delta_{nm}$ is the Kronecker delta.
Let us introduce a trial eigenstate $| q_{\gamma} \rangle$ with the corresponding eigenvalue
$q_{\gamma}$, which satisfies
\begin{align}
\sum_m [ H^{[ n - m]} + n \hbar \omega \delta_{nm} ] | q_{\gamma} \rangle
= q_{\gamma} | q_{\gamma} \rangle,
\end{align}
and the integers $m$ and $n$ run from $- \infty$ to $\infty$.
Using orthogonal states labeled by $\alpha$ and $\beta$ in the two-level system,
Eq.~(\ref{eqn:Hq}) is rewritten as
\begin{align}
\sum_m \sum_\beta \langle \alpha n | H_F | \beta m \rangle
\langle \beta m | q_{\gamma} \rangle = q_{\gamma} \langle \alpha n | q_{\gamma} \rangle.
\label{eqn:HF}
\end{align}
Here, the eigenvector
$\langle \alpha n | q_{\gamma} \rangle$ ($\equiv \phi^{[n]}_{\alpha \gamma}$) is
represented by a Floquet state $| \alpha n \rangle = | \alpha \rangle \otimes | n \rangle$ which is
considered as an oscillating-field dressed state ($| n \rangle$ is also related to
$\langle t | n \rangle \equiv e^{i n \omega t}$).
The matrix expression of the Floquet Hamiltonian
\begin{align}
\langle \alpha n | H_F | \beta m \rangle = H_{\alpha \beta}^{[n - m]}
+ n \hbar \omega \delta_{nm} \delta_{\alpha \beta}
\label{eqn:HFmat}
\end{align}
is constructed by an infinite number of $| \alpha n \rangle$.
Indeed, $H_F | q_{\gamma} \rangle = q_{\gamma} | q_{\gamma} \rangle$ in Eq.~(\ref{eqn:HF}) is
extended to the generalized eigenvalue problem
$H_F | q_{\gamma l} \rangle = q_{\gamma l} | q_{\gamma l} \rangle$, where 
$q_{\gamma l} = q_{\gamma} + l \hbar \omega$ and
$\langle \alpha n | q_{\gamma l} \rangle = \langle \alpha, n - l | q_{\gamma} \rangle$ are satisfied.
\cite{Shirley65,Nakai16}
Owing to this translational invariance, it is sufficient to solve the essential eigenvalue
$q_{\gamma}$ by diagonalizing $H_F$.
In the following discussion, $\hbar = 1$ is used for simplicity.
\par

Let us apply the above matrix expression of the Floquet Hamiltonian in Eq.~(\ref{eqn:HFmat}) to
the two-level system coupled to the SAW and the MW in Eq.~(\ref{eqn:Ht}), where the SAW
phonon frequency $\omega_m$ is replaced by $\omega$.
First, the block Hamiltonian in the subspace of $\{ | g n \rangle, | e n \rangle \}$ for the same
integer $n$ can be written as
\begin{align}
H_n^{[0]} \equiv H^{[0]} + n \omega =
\left(
\begin{array}{cc}
- ( \varepsilon_0 / 2 ) + n \omega & \Delta^* / 2 \\
\Delta / 2 & ( \varepsilon_0 / 2 ) + n \omega
\end{array}
\right).
\label{eqn:Hn0}
\end{align}
There are only finite matrix elements between $| g~({\rm or}~e), n \rangle$ and
$| g~({\rm or}~e), m = n \pm 1 \rangle$ in Eq.~(\ref{eqn:HFmat}), and the corresponding block
matrix parts are given by
\begin{align}
H^{[1]} =
\frac{1}{4} \left(
\begin{array}{cc}
- A_L & A_T \\
A_T & A_L
\end{array}
\right),~~
H^{[-1]} =
\frac{1}{4} \left(
\begin{array}{cc}
- A_L^* & A_T^* \\
A_T^* & A_L^*
\end{array}
\right),
\end{align}
while $H^{[n]} = {\bf 0}$ ($|n| \ge 2)$, where all elements are zeros in the $2 \times 2$ matrix
$\bf 0$.
Using these block matrices, the Floquet matrix is written as
\begin{align}
H_F =
\left(
\begin{array}{ccccccc}
\ddots & ~ & ~ & \vdots & ~ & ~ & ~ \\
~ & H_{-2}^{[0]} & H^{[-1]} & {\bf 0} & {\bf 0} & {\bf 0} & ~ \\
~ & H^{[1]} & H_{-1}^{[0]} & H^{[-1]} & {\bf 0} & {\bf 0} & ~ \\
\cdots & {\bf 0} & H^{[1]} & H_0^{[0]} & H^{[-1]} & {\bf 0} & \cdots \\
~ & {\bf 0} & {\bf 0} & H^{[1]} & H_1^{[0]} & H^{[-1]} & ~ \\
~ & {\bf 0} & {\bf 0} & {\bf 0} & H^{[1]} & H_2^{[0]} & ~ \\
~ & ~ & ~ & \vdots & ~ & ~ & \ddots
\end{array}
\right).
\label{eqn:HFblock}
\end{align}
Eigenvalues and the corresponding eigenvectors are solved by a finite number of blocks of the
Floquet matrix.
As reported by Ref.~16,
we choose $n \simeq 50$, namely, from $H_{-50}^{[0]}$ to $H_{50}^{[0]}$ in Eq.~(\ref{eqn:HFblock}), and check that the numerical results are sufficiently converged.
Here, we devote ourselves to calculate the time-averaged transition probability between the
two states $| g \rangle$ and $| e \rangle$,
\cite{Shirley65}
\begin{align}
\bar{P}_{g \rightarrow e} =
\sum_m \sum_{\gamma} | \langle e m | q_{\gamma} \rangle \langle q_{\gamma} | g 0 \rangle |^2,
\label{eqn:TP}
\end{align}
in the energy region $0 < \varepsilon_0 < 2 \omega$, and focus on a single phonon process at
around $\varepsilon_0 = \omega$.
\par

Inclusion of $T$-mode phonons ($A_T \ne 0$) makes the numerical analysis more complicated
than the case of $A_T = 0$ where only $L$-mode phonons are coupled to photons.
For the latter, sharp peaks of the transition probability are generated at around
$\varepsilon_0 = \omega, 2\omega, \cdots$ for a strong oscillating field of the $L$-mode
$|A_L| / \omega \gg 1$.
Owing to a finite amplitude $A_T$, these peaks become broad and the $n$th peak positions
shift away from $\varepsilon_0 = n \omega$.
It is also marked that the appearance of the transition probability peaks is affected by
the phase difference $\theta_T - \theta_L$ between the two phonon modes
as well as the amplitudes $| A_L | / \omega$ and $| A_T | / \omega$.
In addition, for the photon-phonon coupled process, it is necessary to consider the phase shift
$\phi_l$ of $\Delta = |\Delta| e^{i \phi_l}$ as well as the amplitude $|\Delta|$.
As discussed in the previous subsection, $A_\mu$ ($\mu = L,T$) is related to the
QS coupling amplitude $g_\mu a_\mu$ and the phase shift $\phi_\mu$ of the
$\mu$-mode, which changes with the rotation angle $\theta$ of the magnetic field as shown in
Eq.~(\ref{eqn:ga}).
 
\subsection{Nearly degenerate Floquet states}
It is very useful to consider the nearly degenerate Floquet states which are involved in a
multiphonon process, and this is formulated by the Van Vleck perturbation theory.
The derivation of an effective Hamiltonian for the almost degenerate levels was reported in a
similar case of the two-level system coupled to an oscillating field.
\cite{Son09, Hausinger10}
We extend the previous formulation including a direct off-diagonal coupling of the oscillating field
between the two states, namely, $A_T$ of the $T$-mode phonons in addition to $A_L$ of the
$L$-mode phonons.
\par

Let us start from an unperturbed Hamiltonian $H_{F,0} = H_F (\Delta = 0, A_T = 0)$ in
Eq.~(\ref{eqn:HFblock}), and the eigenvalue equations are given by
\begin{align}
& H_{F,0}~ | \tilde{g} n \rangle = \left( - \frac{\varepsilon_0}{2} + n \omega \right)
| \tilde{g} n \rangle,
\label{eqn:HF0g} \\
& H_{F,0}~ | \tilde{e} m \rangle = \left( \frac{\varepsilon_0}{2} + m \omega \right)
| \tilde{e} m \rangle.
\label{eqn:HF0e}
\end{align}
Equation (\ref{eqn:HF0g}) corresponds to the following original differential equation
\begin{align}
\left\{ \frac{1}{2} [ -\varepsilon_0 - |A_L| \cos ( \omega t + \theta_L ) ]
- i \frac{\partial}{\partial t} \right\} \phi_n (t)
= \lambda \phi_n (t),
\end{align}
where $A_L = | A_L | e^{ i \theta_L }$ and $\lambda =  - (\varepsilon_0 / 2) + n \omega$.
The analytic solution is obtained as
\begin{align}
& \phi_n (t) = e^{ i n \omega t} e^{ i [|A_L| / (2 \omega)] \sin ( \omega t + \theta_L )}
= \sum_{k = - \infty}^{\infty} e^{ i k \theta_L } J_k \left(  \frac{|A_L|}{2 \omega} \right)
e^{ i (n+k) \omega t }
\nonumber \\
&~~~~~~
= \sum_{k = - \infty}^{\infty} e^{ i (k-n) \theta_L } J_{k-n} \left(  \frac{|A_L|}{2 \omega} \right)
e^{ i k \omega t },
\end{align}
which is expanded with the $k$th order Bessel functions $J_k (z)$.
This leads to the expression of $| \tilde{g} n \rangle$ on the basis of the Floquet states
$| g k \rangle$ as
\begin{align}
| \tilde{g} n \rangle = \sum_{k = - \infty}^{\infty} e^{i(k-n) \theta_L}
J_{k-n} \left( \frac{|A_L|}{2 \omega} \right) | g k \rangle.
\end{align}
In the same manner for Eq.~(\ref{eqn:HF0e}),
\begin{align}
| \tilde{e} m \rangle = \sum_{k = - \infty}^{\infty} e^{i(k-m) \theta_L}
J_{k-m} \left( - \frac{|A_L|}{2 \omega} \right) | e k \rangle.
\end{align}
\par

Next, for finite $\Delta$ and $A_T$, the Floquet matrix elements are rewritten on the new basis set
of $\{ | \tilde{g} n \rangle, | \tilde{e} m \rangle \}$ for any integer $n$ and $m$.
Using $J_n (2z) = \sum_m J_{n-m} (z) J_m (z)$ for $z = | A_L | / 2 \omega$, the matrix element
related to the transition $e \rightarrow g$ is calculated as 
\begin{align}
& \langle \tilde{g} n | H_F | \tilde{e} m \rangle
= \sum_{k = -\infty}^{\infty} \sum_{l = -\infty}^{\infty} e^{ - i (k-n) \theta_L } J_{k-n} (z) J_{l-m} (-z)
\nonumber \\
&~~~~~~~~~~~~~~~~~~~~~~~~
\times
e^{ i (l-m) \theta_L } \langle g k | H_F | e l \rangle
\nonumber \\
&~~~~~~~~~~~~~~~~~~
= \frac{ \Delta^* }{2} \sum_{k = -\infty}^{\infty} e^{ i (n-m) \theta_L } J_{k-n} (z) J_{-k+m} (z)
\nonumber \\
&~~~~~~~~~~~~~~~~~~~~~~
+ \frac{ | A_T | }{4} e^{ i \theta_T } \sum_{k = -\infty}^{\infty} e^{ i (n-m-1) \theta_L }
J_{k-n} (z) J_{-k+m+1} (z)
\nonumber \\
&~~~~~~~~~~~~~~~~~~~~~~
+ \frac{ | A_T | }{4} e^{ - i \theta_T } \sum_{k = -\infty}^{\infty} e^{ i (n-m+1) \theta_L }
J_{k-n} (z) J_{-k+m-1} (z)
\nonumber \\
&~~~~~~~~~~~~~~~~~~
= e^{ -i (m-n) \theta_L } \left[ \frac{ \Delta^* }{2} J_{m-n} (2z)
+ \frac{ | A_T | }{4} e^{ i ( \theta_T - \theta_L ) } J_{m-n+1} (2z)
\right.
\nonumber \\
&~~~~~~~~~~~~~~~~~~~~~~\left.
+ \frac{ | A_T | }{4} e^{ - i ( \theta_T - \theta_L ) } J_{m-n-1} (2z) \right]
\equiv v_{m-n}^{\tilde{g} \tilde{e}}.
\label{eqn:vge}
\end{align}
In the same manner,
$\langle \tilde{e} n | H_F | \tilde{g} m \rangle \equiv v_{n-m}^{\tilde{e} \tilde{g}}
= ( v_{n-m}^{\tilde{g} \tilde{e}} )^*$ is obtained for the transition $g \rightarrow e$.
For the matrix element related to the transition $g \rightarrow g$,
\begin{align}
& \langle \tilde{g} n | H_F | \tilde{g} m \rangle
= \sum_{k = -\infty}^{\infty} \sum_{l = -\infty}^{\infty} e^{ - i (k-n) \theta_L } J_{k-n} (z) J_{l-m} (z)
\nonumber \\
&~~~~~~~~~~~~~~~~~~~~~~~~
\times e^{ i (l-m) \theta_L } \langle g k | H_F | g l \rangle
\nonumber \\
&~~~~~~~~~~~~~~~~~~
= \sum_{k = -\infty}^{\infty} e^{ i (n-m) \theta_L } J_{k-n} (z)
\left\{ \left( - \frac{ \varepsilon_0 }{2} + k \omega \right) J_{k-m} (z)
\right.
\nonumber \\
&~~~~~~~~~~~~~~~~~~~~~~\left.
- \frac{ | A_L | }{4} \left[ J_{k-m-1} (z) + J_{k-m+1} (z) \right] \right\}.
\end{align}
Using $J_{n-1} (z) + J_{n+1} (z) = 2nJ_n (z) / z$ and the summation
$\sum_m J_m (z) J_{m-n} (z) = \delta_{n0}$, we obtain
\begin{align}
& \langle \tilde{g} n | H_F | \tilde{g} m \rangle
= e^{ i (n-m) \theta_L } \left( - \frac{ \varepsilon_0 }{2} + m \omega \right)
\sum_{k = -\infty}^{\infty} J_{k-n} (z) J_{k-m} (z)
\nonumber \\
&~~~~~~~~~~~~~~~~~~
= \left( - \frac{ \varepsilon_0 }{2} + n \omega \right) \delta_{nm}.
\end{align}
In the same manner, for the transition $e \rightarrow e$,
\begin{align}
\langle \tilde{e} n | H_F | \tilde{e} m \rangle
= \left( \frac{ \varepsilon_0 }{2} + n \omega \right) \delta_{nm}.
\end{align}
In the above calculations, the following matrix elements have been used:  
\begin{align}
& \langle g k | H_F | e l \rangle = \frac{\Delta^*}{2} \delta_{kl}
+ \frac{ | A_T | }{4} ( e^{ i \theta_T } \delta_{k, l+1} + e^{ - i \theta_T } \delta_{k, l-1} ), \\
& \langle e k | H_F | g l \rangle = \frac{\Delta}{2} \delta_{kl}
+ \frac{ | A_T | }{4} ( e^{ i \theta_T } \delta_{k, l+1} + e^{ - i \theta_T } \delta_{k, l-1} ), \\
& \langle g k | H_F | g l \rangle = \left( - \frac{ \varepsilon_0 }{2} + k \omega \right) \delta_{kl}
- \frac{ | A_L | }{4} ( e^{ i \theta_L } \delta_{k, l+1} + e^{ - i \theta_L } \delta_{k, l-1} ), \\
& \langle e k | H_F | e l \rangle = \left( \frac{ \varepsilon_0 }{2} + k \omega \right) \delta_{kl}
+ \frac{ | A_L | }{4} ( e^{ i \theta_L } \delta_{k, l+1} + e^{ - i \theta_L } \delta_{k, l-1} ).
\end{align}
\par

Let us consider that the nearly degenerate states $| \tilde{g} 0 \rangle$ and $| \tilde{e}, -n \rangle$
are coupled to each other through $v_{-n}^{\tilde{g} \tilde{e}}$ defined in Eq.~(\ref{eqn:vge}),
where $- \varepsilon_0 / 2 \simeq \varepsilon_0 / 2 - n \omega$.
The infinite-dimensional matrix of $H_F$ on the basis of
$\{ | \tilde{g} n \rangle, | \tilde{e} m \rangle \}$ is reduced to an effective $2 \times 2$ matrix within
$| \tilde{g} 0 \rangle$ and $| \tilde{e}, -n \rangle$ using the Van Vleck perturbation theory.
The perturbation parameters are $| \Delta |$ and $| A_T |$.
Using a similar matrix form in Refs.~16 and 17, we present an effective Hamiltonian for
$| \Delta | / \omega, | A_T | / \omega \ll 1$:
\begin{align}
\tilde{H}_F =
\left(
\begin{array}{cc}
- \ds{\frac{ \varepsilon_0 }{2} + \delta} & v_{-n}^{\tilde{g} \tilde{e}} \\
v_{-n}^{\tilde{e} \tilde{g}} & \ds{\frac{ \varepsilon_0 }{2} - \delta - n} \omega
\end{array}
\right),
\end{align}
where $\delta$ is the energy shift that corresponds to the ac Stark shift,
\cite{Son09,Hausinger10}
and the off-diagonal matrix
element is represented by the first order of the perturbation parameters as
\begin{align}
& v_k^{\tilde{g} \tilde{e}} = \left( v_k^{\tilde{e} \tilde{g}} \right)^*
\nonumber \\
&~~~~
= e^{ - i k \theta_L } \left[ \frac{\Delta^*}{2} J_k \left( \frac{ | A_L | }{\omega} \right)
+ \frac{ | A_T | }{4} e^{ i ( \theta_T - \theta_L ) } J_{k+1}  \left( \frac{ | A_L | }{\omega} \right)
\right.
\nonumber \\
&~~~~~~~~\left.
+ \frac{ | A_T | }{4} e^{ - i ( \theta_T - \theta_L ) } J_{k-1}  \left( \frac{ | A_L | }{\omega} \right) \right].
\end{align}
The leading term of $\delta$ is given by
\begin{align}
\delta = - \sum_{~_{k \ne -n}^{k = -\infty}}^{\infty}
\frac{ | v_k^{\tilde{g} \tilde{e}} |^2 }{ \varepsilon_0 + k \omega },
\label{eqn:shift}
\end{align}
which indicates that an $n$-phonon resonance occurs at $\varepsilon_0 = n \omega + 2 \delta$.
Indeed, the eigenvalues of $\tilde{H}_F$ are
\begin{align}
q_{\pm} = - \frac{n \omega}{2}
\pm \sqrt{ \frac{ ( n \omega - \varepsilon_0 + 2 \delta )^2 }{4} + | v_{-n}^{\tilde{g} \tilde{e}} |^2 },
\end{align}
which lead to the time-dependent transition probability from $| g \rangle$ to
$| e \rangle$  mediated by $n$ phonons
\begin{align}
P_{g \rightarrow e}^{(n)} (t) = \frac{ | v_{-n}^{\tilde{g} \tilde{e}} |^2 }{ \tilde{q}^2 }
\sin^2 \tilde{q} t
= \frac{ | v_{-n}^{\tilde{g} \tilde{e}} |^2 }{ \tilde{q}^2 } \frac{ 1 - \cos 2 \tilde{q} t }{2},
\end{align}
where
$\tilde{q} = \sqrt{ ( n \omega - \varepsilon_0 + 2 \delta )^2 / 4 + | v_{-n}^{\tilde{g} \tilde{e}} |^2 }$.
In the long-time limit, we obtain the $n$-phonon time-averaged transition probability
\begin{align}
\bar{P}_{g \rightarrow e}^{(n)}
= \frac{1}{2} \frac{ | v_{-n}^{\tilde{g} \tilde{e}} |^2 }
{ | v_{-n}^{\tilde{g} \tilde{e}} |^2 + ( n \omega - \varepsilon_0 + 2 \delta )^2 / 4 }.
\label{eqn:Pn}
\end{align}
The broadening of the transition peaks $4 | v_{-n}^{\tilde{g} \tilde{e}} |$ depends on the following
model parameters
\begin{align}
\Delta \equiv | \Delta |^{i \phi_l},~~\theta_{TL} \equiv \theta_T - \theta_L,~~
J_k \equiv J_k \left( \frac{| A_L |}{\omega} \right),
\end{align}
and is explicitly written as
\begin{align}
& | v_{-n}^{\tilde{g} \tilde{e}} |^2
= \frac{1}{4} \left\{ J_{-n}^2
\left( | \Delta | \cos \phi_l - n \omega \frac{ | A_T | }{ | A_L | } \cos \theta_{TL} \right)^2
\right.
\nonumber \\
&~~~~~~~~~~~~\left.
+ \bigg[ J_{-n} | \Delta | \sin \phi_l \right.
\nonumber \\
&~~~~~~~~~~~~~~~~~~\left.
- \frac{ | A_T | }{2} ( J_{-n+1} - J_{-n-1} ) \sin \theta_{TL} \bigg]^2
\right\}.
\label{eqn:broad}
\end{align} 
\par

\section{Results}
First, the $A_L$ and $A_T$ dependence of the transition probability in the two-level system is
investigated by solving Eqs.~(\ref{eqn:Hn0})--(\ref{eqn:HFblock}).
When $| A_L |$ and $| A_T |$ are fixed, the transition probability depends on the phase
difference $\theta_{TL}$ between the $L$-mode and $T$-mode phonons.
Although it is also dependent on the photon coupling represented by $| \Delta | e^{i \phi_l}$,
the phase $\phi_l$ effect on the transition probability is negligibly small for
$| \Delta | / \omega \ll 1$, where $\omega$ is the energy of a single phonon.
\par

Next, the roles of QS couplings in the transition probability are elucidated for
the two basis states split by a magnetic field and coupled to each other through the quadrupole
operators in Eqs.~(\ref{eqn:Ou2}) and (\ref{eqn:Ozx2}).
As described in Sect.~2.2, $\omega_0$ corresponds to the Zeeman energy of the level splitting.
The QS couplings $A_L$ and $A_T$ in Eq.~(\ref{eqn:parameter}) depend on the
rotation angle $\theta$ of the magnetic field.
Both couplings are determined from the combination of the three strain modes in
Eq.~(\ref{eqn:ga}).
For a quartet system such as the $\Gamma_8$ state in the silicon vacancy, we have
assumed that the magnetic field is strong enough ($\varepsilon_0 / \omega \sim 1$) to neglect the
contribution from the higher excited states, namely, the $O_v$ coupling between the ground and
second excited states.
Although we focus on a single phonon process near $\varepsilon_0 / \omega = 1$, we
show the results for the two-level system including the $\varepsilon_0 / \omega \ll 1$ parameter
region to elucidate the roles of longitudinal and transverse QS couplings in the field-dependent
transition probability.
Indeed, inclusion of the higher excited states in the present model is required for a quantitative
analysis of the quartet case, especially for a small magnetic field.

\subsection{Effects of longitudinal and transverse phonon couplings on time-averaged
transition probability}
\begin{figure}
\begin{center}
\includegraphics[width=7cm,clip]{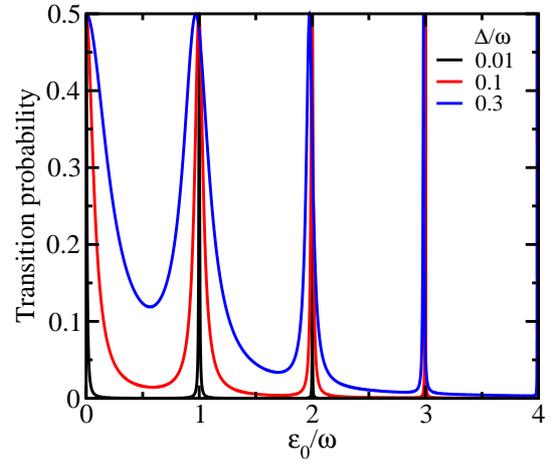}
\end{center}
\caption{
(Color online)
Transition probability for the $L$-mode phonons plotted as a function of the
energy difference between the two states $\varepsilon_0 / \omega$ for various values of the
photon coupling $\Delta$.
Here, $\omega$ is the phonon energy that corresponds to the SAW frequency.
$A_L / \omega = 1$ is chosen for the $L$-mode coupling and $A_T = 0$.
}
\label{fig:3}
\end{figure}
\begin{figure}
\begin{center}
\includegraphics[width=7cm,clip]{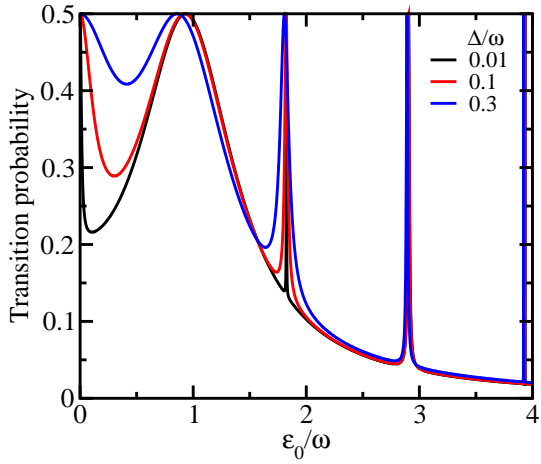}
\end{center}
\caption{
(Color online)
Transition probability for the $T$-mode phonons plotted as a function of $\varepsilon_0 / \omega$
for various $\Delta$, where $A_T / \omega = 1$ is chosen for the $T$-mode coupling and
$A_L = 0$.
}
\label{fig:4}
\end{figure}
\begin{figure}
\begin{center}
\includegraphics[width=7cm,clip]{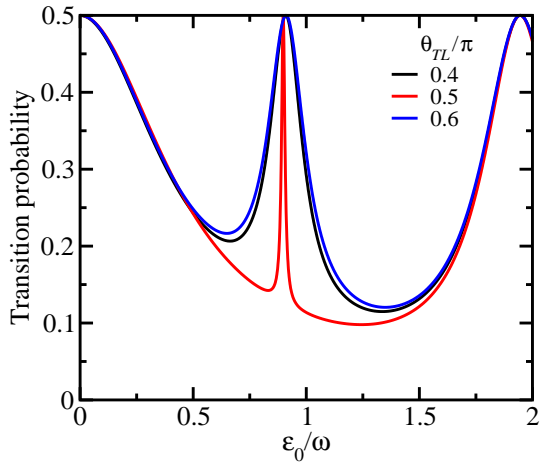}
\end{center}
\caption{
(Color online)
Transition probability for the two-level system coupled to both $L$-mode and $T$-mode phonons
plotted as a function of $\varepsilon_0 / \omega$ for $\Delta / \omega = 0.01$.
The amplitudes of the $L$-mode ($A_L = | A_L | e^{i \theta_L}$) and $T$-mode
($A_T = | A_T | e^{i \theta_T}$) couplings are fixed as $| A_L | / \omega = 1.8$ and
$| A_T | / \omega = 0.8$.
The broadness of the peak near $\varepsilon_0 / \omega = 1$ is sensitively dependent on
the phase difference of the two modes $\theta_{TL} = \theta_T - \theta_L$ around
$\theta_{TL} / \pi = 0.5$.
}
\label{fig:5}
\end{figure}
Let us consider the $L$-mode for $A_T = 0$ at first.
The detailed study was already reported by a similar model used for the multiphoton quantum
interference.
\cite{Son09}
Here, the role of photons is replaced by that of $L$-mode phonons, while the photon is involved in
the transition probability at $\varepsilon_0 = 0$ if there is no contribution from the phonons.
For $A_L = 0$, the transition probability shows the maximum $1/2$ only at $\varepsilon_0 = 0$
that corresponds to a low magnetic field $H_z \sim \omega_l \ll \omega$.
On the other hand, in the absence of the photon coupling ($\Delta = 0$), the transition
cannot be caused by the $L$-mode phonons for $A_L \ne 0$.
Thus, the photon-assisted process is required for a finite transition probability coupled to the
$L$-mode phonons, and the transition probability peaks reach $1/2$ at nearly
$\varepsilon_0 = \omega, 2 \omega, \cdots$.
This indicates a possibility of MAR with the use of the $L$-mode phonons.
Figure~\ref{fig:3} shows that the transition probability peaks become sharp at
$\varepsilon_0 = n \omega$ for $| \Delta | / \omega \ll 1$.
As $| \Delta |$ increases, each peak broadens and the $n$th peak position $\varepsilon_0$
shifts towards a lower value from $\varepsilon_0 = n \omega$.
In particular, the single phonon process at $\varepsilon_0 = \omega$ shows the
largest shift.
This result is inferred from Eqs.~(\ref{eqn:shift}), (\ref{eqn:Pn}) and (\ref{eqn:broad}).
The broadness of the peaks are represented approximately by
$4 | v_{-n}^{\tilde{g} \tilde{e}} | \simeq 2 | \Delta | J_n ( | A_L | / \omega )$ and the peak shifts
also obey $| v_{k \ne -n}^{\tilde{g} \tilde{e}} |^2$ for $| \Delta | / \omega \ll 1$.
The calculated transition probability for $A_T = 0$ does not depend on the phase shifts $\phi_l$
and $\theta_L$.
\par

Next, we show the role of the $T$-mode phonon in the absence of the $L$-mode for $A_L = 0$.
This was originally investigated by Shirley, formulating the Floquet theory for the two-level system
with an oscillating field, and corresponds to $\Delta \rightarrow 0$ in our model.
\cite{Shirley65}
In particular, for $| A_T | / \omega \lesssim 1$, the transition probability shows narrow peaks
near $\varepsilon_0 = 3 \omega, 5 \omega, \cdots$ and a relatively broad peak near
$\varepsilon_0 = \omega$.
When $| \Delta |$ is finite, additional narrow peaks emerge near
$\varepsilon_0 = 2 \omega, 4 \omega, \cdots$ and the peak widths become broader with
increasing $| \Delta |$ as shown in Fig.~\ref{fig:4}.
For a relatively large $| \Delta |$, the peak at $\varepsilon_0 \simeq \omega$ merges into the peak
at $\varepsilon_0 = 0$, and the transition probability shows an almost constant value close to
$1/2$ in $0 < \varepsilon_0 < \omega$.
This indicates that the contribution from the higher excited states may not be
negligible in the quartet case discussed in Sect.~2.2.
\par

To elucidate the roles of the phonon coupling with the $L$- and $T$-modes together in the
transition probability, we devote ourselves to a small $| \Delta | / \omega < 0.1$ for the photon
coupling, where the effect of the photon phase shift $\phi_l$ is negligible.
When both $A_L$ and $A_T$ are finite, the phase difference $\theta_{TL}$ of the two phonon
modes affects the transition probability to some extent.
This is an important point for the MAR assisted by photons.
Here, we examine a single phonon resonance for $| A_T | / | A_L | < 1$ to show explicitly a key
role of a finite $A_T$.
The $A_T$ effect brings about the broadness of a narrow peak in the transition probability at
$\varepsilon_0 = \omega$ and also
shifts the peak position $\varepsilon_0$ towards a lower value from $\omega$.
It is specially marked that the sharpness of the peak becomes prominent for
$| A_L | / \omega \simeq 1.8$.
This is explained by Eq.~(\ref{eqn:broad}) in which the difference of the Bessel functions
$( J_0 - J_2 )$ almost vanishes for $n = 1$.
In addition, the resonance peak becomes narrower when $\theta_{TL}$ is adjusted to satisfy
$\cos \theta_{TL} = (| \Delta | | A_L |) / (\omega | A_T |) \simeq 1.8 | \Delta | / | A_T | < 1$, where
$| v_{-1}^{\tilde{g} \tilde{e}} |$ is almost close to zero.
The $\theta_{TL}$ dependence is shown for $\Delta / \omega = 0.01$,
$| A_L | / \omega = 1.8$ and $| A_T| / \omega = 0.8$ in Fig.~\ref{fig:5}.
According to Eq.~(\ref{eqn:ga}), the QS couplings depend on the rotation angle $\theta$ of the
magnetic field.
Thus, $A_L = - g_L a_L e^{i \phi_L}$ and $A_T = g_T a_T e^{i \phi_T}$ can be controlled
by the field angle for the appearance of such a sharp transition probability peak as
demonstrated in the next subsection.

\subsection{Effects of quadrupole-strain couplings on time-averaged transition probability}
\begin{figure}
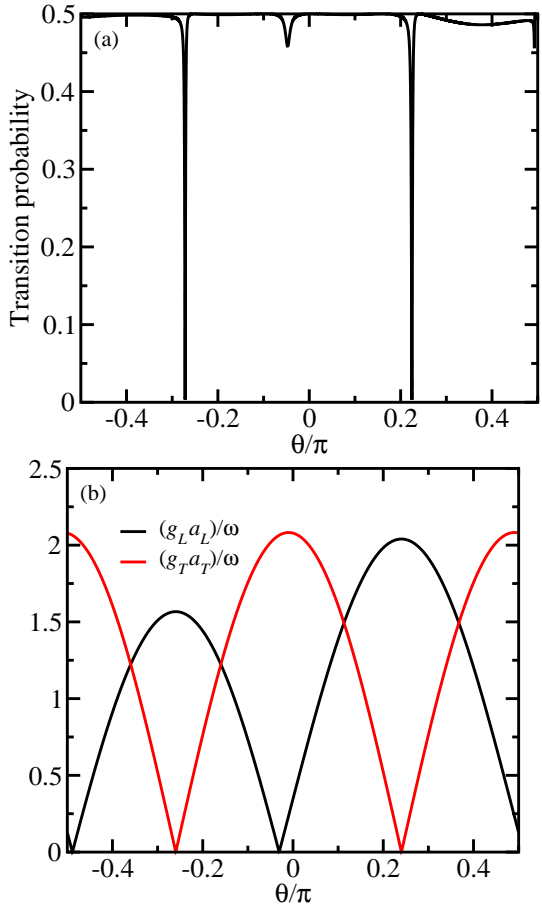

\begin{center}
\includegraphics[width=7cm,clip]{fig6a.eps}
\includegraphics[width=6.5cm,clip]{fig6b.eps}
\end{center}
\caption{
(Color online)
(a) Transition probability $P_{\varepsilon_0 = \omega}$ at $\varepsilon_0 / \omega = 1$
plotted as a function of the magnetic-field angle $\theta$ for $\Delta / \omega = 0.1$.
(b) $L$-mode ($g_L a_L$) and $T$-mode ($g_T a_T$) QS couplings plotted as
a function of $\theta$.
The data are plotted for $(g_{\Gamma_3} a_U) / \omega = 0.1$,
$(g_{\Gamma_5} a_{ZX}) / \omega = 0.6$, and $\phi_U = \phi_V = \phi_{ZX} = 0$.
}
\label{fig:6}
\end{figure}
\begin{figure}
\begin{center}
\includegraphics[width=7cm,clip]{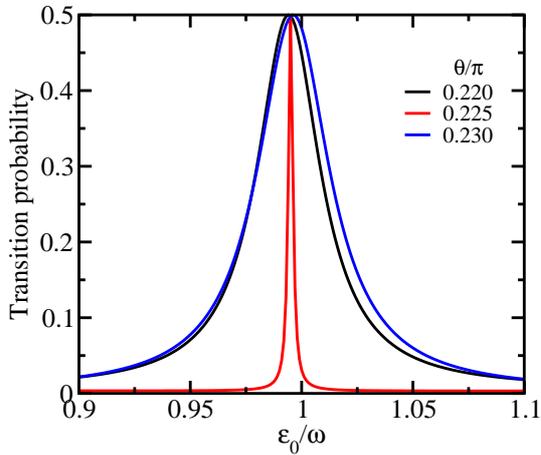}
\end{center}
\caption{
(Color online)
Transition probability for single phonon process.
The data are plotted as a function of
$\varepsilon_0 / \omega$ for various values of $\theta$ near $\theta / \pi = 0.225$ at which
$P_{\varepsilon_0 = \omega}$ shows the narrow dip in Fig.~\ref{fig:6}.
}
\label{fig:7}
\end{figure}
On the basis of the QS couplings described by
Eqs.~(\ref{eqn:HQSLT})--(\ref{eqn:CT2}), we change the quadrupole coupling constants
($g_{\Gamma_3}$ and $g_{\Gamma_5}$), the local strain amplitudes
($a_U$, $a_V$, and $a_{ZX}$) and the relative time-independent phase shifts of the SAW modes
($\phi_U$ and $\phi_V$ measured from $\phi_{ZX} = 0$).
Here, $a_U = a_V$ is set for simplicity, so that the weight of the $\Gamma_3$ and $\Gamma_5$
QS couplings in the two-level system is represented by the ratio of
$g_{\Gamma_3} a_U$ and $g_{\Gamma_5} a_{ZX}$.
The energy difference between the two states is changed as $\varepsilon_0 \propto H_z$ and the
$z$ axis of the field orientation is rotated as $( \sin \theta, 0, \cos \theta )$ in the $ZX$ plane
of the crystal (see Fig.~\ref{fig:2}).
First, we compute $| A_L |$, $| A_T |$ and $\theta_{TL}$ as a function of the field angle $\theta$ for
given $g_{\Gamma_3} a_U$, $g_{\Gamma_5} a_{ZX}$, $\phi_U$, and $\phi_V$.
Next, we solve the eigenvalue problem of $H_F$ in Eq.~(\ref{eqn:HFblock}) to obtain the
time-averaged transition probability in Eq.~(\ref{eqn:TP}) as a function of $\theta$ for the fixed
$\varepsilon_0 = \omega$ and as a function of $\varepsilon_0$ for various values of $\theta$.
In the following, a single phonon process is intensively studied to elucidate a new aspect of
the quadrupoles.
\par

Figure~\ref{fig:6} (a) shows a $\Gamma_5$-component dominant case
($g_{\Gamma_5} a_{ZX}) > (g_{\Gamma_3} a_U)$, choosing
$\phi_U = \phi_V = \phi_{ZX} = 0$ and fixing the two-level splitting at
$\varepsilon_0 = \omega$.
The transition probability is plotted as a function of the field angle $\theta$, and it shows very
narrow dips at $\theta \simeq \pm \pi / 4$.
In Eq.~(\ref{eqn:ga}), $| A_T | = g_T a_T \simeq | C_{T,1} \cos 2 \theta |$ owing to
$| C_{T,1} | \gg | C_{T,2} |$, and $| A_T |$ almost vanishes at $\theta \simeq \pm \pi / 4$,
namely, $| A_T | \ll | A_L | = g_L a_L$.
Both $g_L a_L$ and $g_T a_T$ are also plotted as a function of $\theta$ in Fig.~\ref{fig:6}
(b) and indicates that the $L$-mode coupling is dominant at
$\theta \simeq \pm \pi / 4$.
The origin of the dips is explained by the appearance of a sharp transition probability peak
as shown in Fig.~\ref{fig:7}.
Notice that the energy of the peak position shifts towards a lower value from
$\varepsilon_0 = \omega$.
When $\theta$ is varied from the above value for the sharp peak ($\theta / \pi = 0.225$ in
Fig.~\ref{fig:7}), the peak broadening causes the abrupt increase in the $\theta$-dependent
transition probability $P_{\varepsilon_0 = \omega}$ at $\varepsilon_0 = \omega$ in
Fig.~\ref{fig:6} (a).
This is a key to the microscopic measurement of the quadrupole components.
The $L$-mode-phonon-mediated transition is always accompanied by the photon
absorption, and the corresponding resonance can be detected by the EPR measurement.
For the same phase shifts $\phi_U = \phi_V = \phi_{ZX}$, the narrow dips in the
$\theta$-dependent $P_{\varepsilon_0 = \omega}$ become invisible as the photon coupling
$| \Delta |$ decreases.
It is because the resonance peak appears exactly at $\varepsilon_0 = \omega$.
Instead, the similar narrow dips can be found when $\varepsilon_0$ is slightly shifted from
$\omega$ (for instance, $\varepsilon_0 / \omega = 1.01$).
\par

\begin{figure}
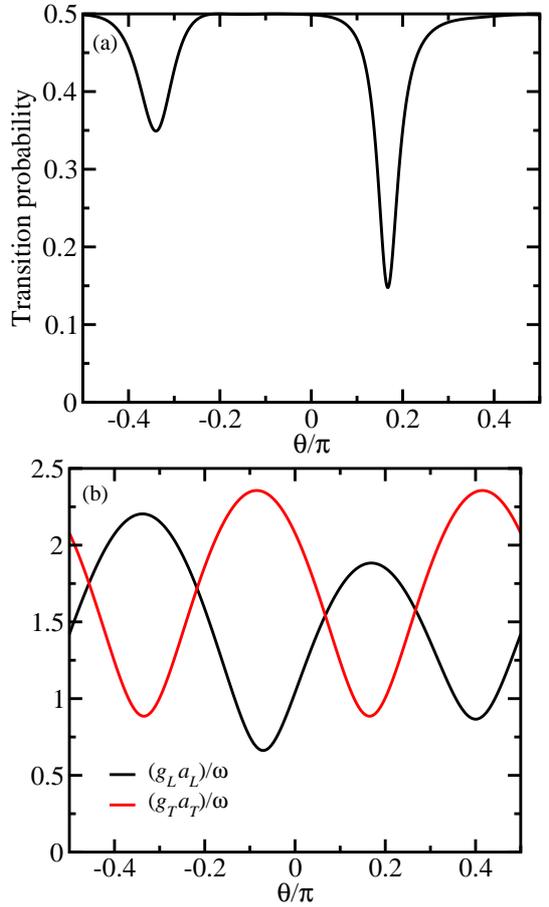

\begin{center}
\includegraphics[width=7cm,clip]{fig8a.eps}
\includegraphics[width=6.5cm,clip]{fig8b.eps}
\end{center}
\caption{
(Color online)
(a) Transition probability $P_{\varepsilon_0 = \omega}$ at $\varepsilon_0 / \omega = 1$
plotted as a function of the magnetic-field angle $\theta$ for $\Delta / \omega = 0.01$.
(b) $L$-mode ($g_L a_L$) and $T$-mode ($g_T a_T$) QS couplings plotted as
a function of $\theta$.
The data are plotted for $(g_{\Gamma_3} a_U) / \omega = 0.3$,
$(g_{\Gamma_5} a_{ZX}) / \omega = 0.6$, and the phase shifts of the three SAW strain modes are
chosen as $\phi_U / \pi = 0.25$, $\phi_V / \pi = - 0.75$, and $\phi_{ZX} = 0$.
}
\label{fig:8}
\end{figure}
\begin{figure}
\begin{center}
\includegraphics[width=7cm,clip]{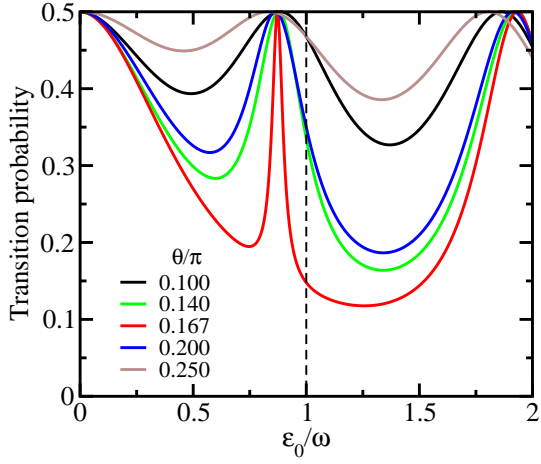}
\end{center}
\caption{
(Color online)
Transition probability plotted as a function of $\varepsilon_0 / \omega$ for various $\theta$
around $\theta / \pi = 0.167$ for the minimum $P_{\varepsilon_0 = \omega}$ in Fig.~\ref{fig:8}.
}
\label{fig:9}
\end{figure}
In generic, the difference of phase shifts $\phi_U$, $\phi_V$, and $\phi_{ZX}$ ($=0$) has to
be considered for the $\theta$ dependence of the transition probability, which can be controlled by
the SAW experiment.
Here, we show the data for $| \phi_U - \phi_V | = \pi$ taking account of the phase difference
between two strains $\varepsilon_U$ and $\varepsilon_V$ reported by the previous SAW
experiment of the silicon wafer.
\cite{Mitsumoto14}
The photon coupling is fixed at $\Delta = 0.01$, which is small enough to treat $\Delta$ as a real
number.
In Fig.~\ref{fig:8} (a), the two dips are also found in the $\theta$-dependent
$P_{\varepsilon_0 = \omega}$.
Since $( g_{\Gamma_5} a_{ZX} ) / ( g_{\Gamma_3} a_U ) > 1$, the field angle
$\theta$ for the minimum transition probability slightly shifts towards a lower value from $\pi / 4$.
This is also explained by the existence of a sharp transition probability peak at a lower value of
$\varepsilon_0$ shifted from $\omega$ as shown in Fig.~\ref{fig:9}.
The resonance near $\varepsilon_0 = \omega$ is owing to the $L$-mode dominant contribution.
In Fig.~\ref{fig:8} (b), $| A_L |$ ($= g_L a_L$) almost reaches a local maximum
$(\simeq 1.8 \omega)$ when $| A_T |$ ($= g_T a_T$) equals a local minimum.
Let us consider the field angle $\theta_{\rm dip}$ for the minimum $P_{\varepsilon_0 = \omega}$.
This $\theta_{\rm dip}$ almost equals the value for the minimum $| A_T |$, which can be evaluated
by Eq.~(\ref{eqn:ga}).
Using $\xi \equiv ( g_{\Gamma_3} a_U ) / ( g_{\Gamma_5} a_{ZX} )$, $| A_T |$ is rewritten as
\begin{align}
\frac{ | A_T |^2 }{ 6 ( g_{\Gamma_5} a_{ZX} )^2 }
= 1 + \xi^2 \left( 1 + \frac{ \sqrt{3} }{2} \right)
+ A_{\phi_U} \cos ( 4 \theta + \alpha ),
\label{eqn:AT2}
\end{align}
and
\begin{align}
& A_{\phi_U}^2 = \left[ 1 - \xi^2 \left( 1 + \frac{ \sqrt{3} }{2} \right) \right]^2
+ \xi^2 ( 1 + \sqrt{3} )^2 \cos^2 \phi_U, \\
& \tan \alpha = \frac{ \xi ( 1 + \sqrt{3} ) }
{ \ds{1 - \xi^2 \left( 1 + \frac{ \sqrt{3} }{2} \right)} } \cos \phi_U.
\label{eqn:phase}
\end{align}
When the $\Gamma_5$ component is dominant as $\xi \ll 1$ and $A_{\phi_U} > 0$, one of the
conditions for the minimum $| A_T |$ is given by $\theta = ( \pi - | \alpha | ) / 4$ for
$0 < | \phi_U | < \pi / 2$.
This equation also indicates that for $\pi / 2 < | \phi_U | < \pi$, the transition probability shows
a similar dip near the negative $\theta = - ( \pi - |\alpha| ) / 4$.
On the other hand, for $\xi \gg 1$ and $A_{\phi_U} < 0$ which correspond to the
$\Gamma_3$-dominant case, one can find that these values $\theta$ are replaced by
$| \alpha | / 4$ and $- | \alpha | / 4$, respectively.
When $\theta$ differs from $\theta_{\rm dip}$ in Fig.~\ref{fig:9}, the transition probability
peak broadens keeping the peak position $\varepsilon_0$ at almost the same value.
This broadening comes from the $T$-mode contribution.
In the quartet case discussed in Sect.~2.2, the contribution from the higher excited
states may not be negligible when the peak at $\varepsilon_0 / \omega \simeq 1$ merges into the
peak at $\varepsilon_0 = 0$ for $\theta / \pi \simeq 0.1$ or $\theta / \pi \simeq 0.25$.
\par

We conclude that it is essential to find the field angle $\theta_{\rm dip}$ for the sharp resonance
peak.
Owing to the photon-assisted transition dominated by the $L$-mode phonon, the resonance
peak can be probed by the EPR measurement.
This $\theta_{\rm dip}$ is related to the ratio $\xi$ of the QS couplings with the
different symmetries $\Gamma_3$ and $\Gamma_5$.
In addition, the ratio of the quadrupole coupling constants $g_{\Gamma_3}$ and
$g_{\Gamma_5}$ can be evaluated from the measurement of the local strain amplitudes
$a_i$ ($i = U, V, ZX$).
\par

\begin{figure}
\begin{center}
\includegraphics[width=7cm,clip]{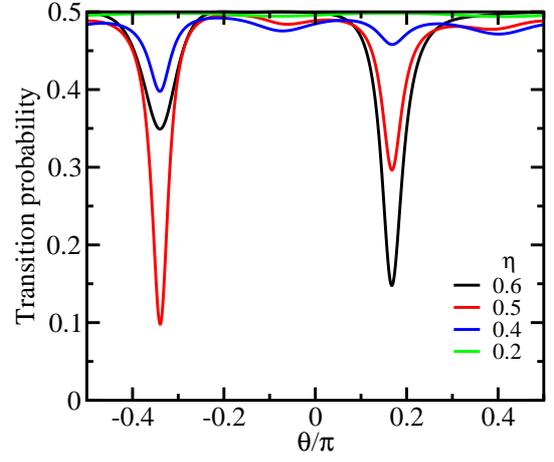}
\end{center}
\caption{
(Color online)
Transition probability $P_{\varepsilon_0 = \omega}$ at $\varepsilon_0 / \omega = 1$
plotted as a function of the magnetic-field angle $\theta$.
The phase shifts of the three SAW strain modes are chosen as $\phi_U / \pi = 0.25$,
$\phi_V / \pi = - 0.75$, and $\phi_{ZX} = 0$.
Here, $g_{\Gamma_3} a_U = g_{\Gamma_3} a_V = 0.5 g_{\Gamma_5} a_{ZX}$, and
the data are plotted for various values of $\eta \equiv (g_{\Gamma_5} a_{ZX}) / \omega \le 0.6$.
The data for $\eta = 0.6$ is the same as $P_{\varepsilon_0 = \omega}$ in Fig.~\ref{fig:8}.
}
\label{fig:10}
\end{figure}
Owing to the penetration of the SAW into the solid, the strain amplitudes $a_i$ show the same
exponential decay along the $Z$ axis
\cite{Mitsumoto14}
and cause a significant decrease in $| A_L |$ and $| A_T |$.
Since the ratio of $g_{\Gamma_3} a_U$, $g_{\Gamma_3} a_V$, and $g_{\Gamma_5} a_{ZX}$
is unchanged, $| A_T | / | A_L |$ keeps the $\theta$ dependence obtained on the surface ($Z = 0$)
during the decay of $a_i$.
Accordingly,  the local minimums of $P_{\varepsilon_0 = \omega}$ appear at almost the same
values $\theta$ as shown in Fig.~\ref{fig:10}, where $\eta$ is defined as the value of
$(g_{\Gamma_5} a_{ZX}) / \omega$ keeping
$g_{\Gamma_3} a_U = g_{\Gamma_3} a_V = 0.5 g_{\Gamma_5} a_{ZX}$.
The origin of the left narrow dip at $\theta / \pi = - 0.340$ for $\eta = 0.5$ is also explained by the
local maximum $| A_L | = g_L a_L \simeq 1.8 \omega$.
This is evaluated from the value of $(5/6) g_L a_L$ in Fig.~\ref{fig:8} (b) plotted for $\eta = 0.6$.
For small $a_i$ ($\eta < 0.2$), $P_{\varepsilon_0 = \omega}$ shows almost constant values
$\simeq 1/2$ in the entire region $- \pi / 2 \le \theta \le \pi / 2$.
Thus, the decay of $a_i$ does not affect the appearance of the dips in the $\theta$-dependent
transition probability.
In the single phonon process at $\varepsilon_0 \simeq \omega$, $| A_L | / \omega \simeq 1.8$
is the most appropriate condition for detecting the photon-assisted $L$-mode phonon resonance,
which is explained by Eq.~(\ref{eqn:broad}) as discussed in Sect.~3.1.
This condition of $| A_L |$ can be accessed by tuning the frequency $\omega$ and the
amplitudes $a_i$ of the SAW strain modes.

\section{Conclusion and Discussion}
In this paper, we have studied a minimum model of the two-level system for the hybrid
EPR-MAR measurement and have demonstrated how the dynamical QS couplings
can be evaluated by the magnetic field angle dependence of the time-averaged transition
probability.
Here, the photon frequency is negligibly small compared to the phonon
frequency of gigahertz order.
The longitudinal ($L$-mode, $A_L$) and transverse ($T$-mode, $A_T$) couplings of the two
states are related to the three phonon modes of elastic strains ($\varepsilon_U$, $\varepsilon_V$,
and $\varepsilon_{ZX}$).
These are driven by the SAW propagating in the principal axis ($X$ axis).
The key point is that both couplings are changed by rotating the magnetic field since the
combination of the three QS couplings depends on the rotation angle.
\par

Using the Floquet theory, we have investigated the effects of $L$-mode and $T$-mode couplings
on the transition probability as a function of $\varepsilon_0 / \omega$, where $\varepsilon_0$ and
$\omega$ are the energy difference of the two states and the SAW phonon frequency,
respectively.
It is essential to focus on the single phonon process for $\varepsilon_0 / \omega \lesssim 1$.
The sharpness of the transition probability peak originates from the $L$-mode coupling
effect assisted by the phonon with a small coupling $\Delta$,
while the peak broadening at $\varepsilon_0 / \omega \simeq 1$ is owing to the $T$-mode
coupling.
The appearance of the transition probability peaks also depends on the phase difference between
the $L$-mode and $T$-mode couplings.
\par

On the basis of this result, we have shown the effects of the QS couplings on
the transition probability at $\varepsilon_0 / \omega = 1$ as a function of the field angle $\theta$
for a $\Gamma_5$-quadrupole dominant case.
The important finding is the appearance of the two narrow dips in the
transition probability owing to the vanishment of $| A_T |$ and the approach of
$| A_L |$ to a local maximum.
The origin of the dips is explained by the lower energy shift of a sharp resonance peak
from $\varepsilon_0 / \omega = 1$ and the strong $\theta$ dependence of the peak
broadening.
Experimentally, the field angle for the sharp resonance can be probed by the EPR measurement
since the photon-assisted transition becomes more significant for $| A_L | > | A_T |$.
This is important for evaluating the ratio of QS couplings with different symmetries.
\par

In the hybrid EPR-MAR measurement as described above, the $L$-mode phonon plays an
important role in the appearance of a sharp resonance peak in the field-angle-dependent
transition probability.
Since the $L$-mode coupling is usually inactive in the transition process, we have introduced
the idea of photon-assisted MAR and demonstrated how to extract a dominant $L$ mode for the
MAR from the various phonon modes.
The sharpness of the resonance peak becomes more prominent for a weak photon filed,
which may allow high sensitivity in the conventional optical measurements.
\par

For the silicon vacancy presented in Sect.~2.2, the ratio of the $\Gamma_3$ and $\Gamma_5$
QS coupling constants $g_{\Gamma_5} / g_{\Gamma_3} = 1.6$ was reported by the
low-temperature ultrasonic measurement.
\cite{Baba13}
This indicates the $\Gamma_5$-quadrupole dominant case as discussed in Sect.~3.2 when the
amplitudes of elastic strains are almost the same in magnitude.
It is expected that the field angle for a sharp resonance appears in $1/8 < \theta / \pi < 1/4$.
Indeed, the $\theta$-dependent transition probability also depends on the phase differences
between the strain modes driven by the SAW.
Thus, the evaluation of the QS coupling constants requires precise measurement of $a_i$ and
$\phi_i$ of the strains represented by $\varepsilon_i = a_i \cos ( \omega t + \phi_i )$
($i = U, V, ZX$).
\par

The idea of the hybrid EPR-MAR measurement can also be applied to reveal hidden quadrupole
properties in the $S = 1$ ground state of the NV center.
For the NV spin state in the $C_{3v}$ crystal-field environment, electronic dipoles belong to
the same point-group character as electric quadrupoles represented by the second-rank spin
tensors:
\cite{Udvarhelyi18,Oort90,Doherty12,Matsumoto17}
$Q_v = S_x^2 - S_y^2$ and $Q_{zx} = S_z S_x + S_x S_z$ for the $x$ component of the dipole;
$Q_{xy} = S_x S_y + S_y S_x$ and $Q_{yz} = S_y S_z + S_z S_y$ for the $y$ component.
Among them, there has been a lack of information on the
relevance of $Q_{zx}$ and $Q_{yz}$ to electric-field control of the NV spin.
\cite{Doherty12,Doherty13}
A possibly significant role of these quadrupoles has been pointed out by the recent theoretical
proposal of mechanically and electrically driven electron spin resonance as a new application of
spin-strain interactions in the NV center.
\cite{Udvarhelyi18}
More detailed investigations of the unknown quadrupole couplings are required to
realize a promising platform of spin-controlled devices for quantum information processing and
sensing applications.
\cite{Suter17}
\par

It will be also intriguing to apply our idea to revisit various quadrupole properties in well-localized
$f$-electron systems such as CeB$_6$.
\cite{Shiina97,Kuramoto09}
It is expected that the antiferroquadrupole ordering transition in CeB$_6$ affects the local
$f$-electron level structure with the $\Gamma_8$ symmetry, which can modify a photon-phonon
coupling process in the EPR measurement under a propagating SAW of gigahertz order.
According to the resonant x-ray diffraction experiment, a linear combination of
quadrupole order parameters is continuously changed by controlling the direction of an applied
magnetic field.
\cite{Matsumura12}
This is a new aspect of quadrupole dynamics that is detectable in various orbitally degenerate
electron systems.

\acknowledgments
This work was supported by JSPS KAKENHI Grant Number 17K05516.

\end{document}